# Building a Vietnamese language query processing framework for e-library searching systems

Dang Tuan Nguyen, Ha Quy-Tinh Luong
Faculty of Computer Science
University of Information Technology, VNU- HCM
Ho Chi Minh city, Vietnam

Tuyen Thi-Thanh Do
Faculty of Software Engineering
University of Information Technology, VNU - HCM
Ho Chi Minh city, Vietnam

*Abstract*—In the objective of building intelligent searching systems for e-libraries or online bookstores, we have proposed a searching system model based on a Vietnamese language query processing component. Such document searching systems based on this model can allow users to use Vietnamese queries that represent content information as input, instead of entering keywords for searching in specific fields in database. To simplify the realization process of system based on this searching system model, we set a target of building a framework to support the rapid development of Vietnamese language query processing components. Such framework let the implementation of Vietnamese language query processing component in similar systems in this domain to be done more easily.

## I. INTRODUCTION

In the objective of building intelligent searching systems for e-libraries or online bookstores, we have proposed a searching system model based on a Vietnamese language query processing component. Such document searching systems based on this model can allow users to use Vietnamese queries that represent content information as input, instead of entering keywords for searching in specific fields in database.

This searching system model includes a restricted parser for analyzing Vietnamese query, a transformer for transforming syntactic structure of query to its semantic representation, a generator for generating queries on relational database from semantic model, and a constructor of answer. In fact, this searching system model inherits the idea of an earlier our document retrieval system, which supports users to use English queries for searching e-books in Gutenberg e-library. [1], [2], [3], [4], [5], [6], [7], [8].

To simplify the realization process of system based on this searching system model, we set a target of building a framework to support the rapid development of Vietnamese language query processing components. Such framework let the implementation of Vietnamese language query processing component in similar systems in this domain to be done more easily.

## II. FRAMEWORK ARCHITECTURE

The VLQP framework is architecture of 2-tiers. This framework includes a restricted parser for analyzing Vietnamese query from users based on a class of the pre-defined syntactic rules and a transformer for transforming syntactic structure of query to its semantic representation.

Main features of those components are described in brief as follows:

- The parser analyzes Vietnamese query syntaxes and output of the syntactic components that were analyzed from the query. After analyzing, the parts-of-speech and the sub-categories of these components are determined. The parser's performing is based on a set of syntactic rules. This set of syntactic rules can cover various forms of Vietnamese query relating to the e-book searching application in e-libraries. The new syntactic rules can be added to the set of these rules for enriching it.

- The transformer bases on predefined transforming rules to transform the syntactic structure of Vietnamese query to its semantic representation. These rules are defined specifically for some determined application domain. The semantic representation model is also built to represent the semantic of all forms of Vietnamese query which are represented by syntactic rules.

The architecture of framework is illustrated in figure 1.

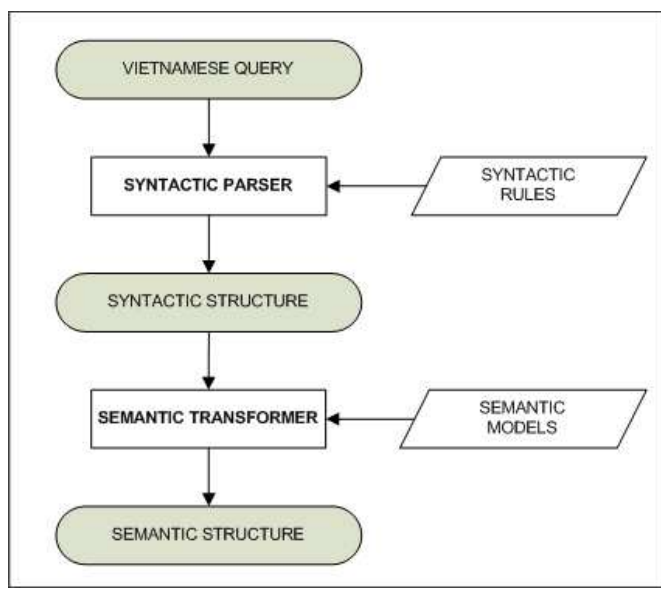

Figure 1. Framework architecture

The VLPQ framework is given as a complete Java package. The Vietnamese language query processing components of searching systems based on VLPQ have an ability of getting Vietnamese queries as input and giving theirs semantic representations as output. The searching systems must build some additional components to process semantic representations of Vietnamese queries and give results to user.

## III. RESTRICTED PARSER

### A. Description of syntactic rules

The parser is built for analyzing the syntax of Vietnamese queries in determined application domain.

For examples, some different query forms as following:

- *Ai đã viết cuốn sách B vào năm 2000?*
  *(Who wrote book B in 2000?)*

- *Nhà xuất bản nào đã phát hành cuốn B trong năm 2008?*

  *(Which publisher published book B in 2008?)*

- *Sách B được tác giả A viết vào năm nào?*
  *(What year did author A write book B?)*

- *Trong năm 2009, tác giả A có viết sách nào thuộc chủ đề T không?*
  *(In 2009, does author A write any book with subject T?)*

The syntax of Vietnamese question forms can be described by BNF notation (Backus–Naur Form). The set of syntactic rules contains about 60 forms of Vietnamese queries involving in titles, authors, years of publication, publishers, subject … For example, the following query's analyzed into syntactic components:

- *S1 := Tác giả A có viết sách B vào năm 2008 không?*
  *(S1:= Did author A write book B in 2008?)*

In this query, the words "có" and "không" are interrogative words. As a result, it can be analyzed into components:

- *author: tác giả A (author A)*
- *interrogative1: có*
- *verb_write: viết (write)*
- *book: sách B (book B)*
- *adverbial phrase of time (APT): vào năm 2008 (in 2008)*
- *interrogative2: không*

The above query is represented in BNF notation:

- *S1_BNF:=<author> [<interrogative1>] <verb> <book> [<APT>] [<interrogative2>] "?"*

### B. Syntactic rules

The parser works on a set of predefined syntactic rules. Table 1 presents a full list of syntactic rules in BNF form which is included in VLPQ framework version 1.0.

TABLE 1. SYNTACTIC RULES

| No | Syntactic rules |
|---|---|
| 1 | <Q1.1a> = <what_author> [<vperfect>] [<interrogative1>] <verb_write> <book> {[<conjunction>] <book>} [<time_phrase>] "?" |
| 2 | <Q1.1b> = [<time_phrase>] [","] <what_author> [<vperfect>] [<interrogative1>] <verb_write> <book> {[<conjunction>] <book>} "?" |
| 3 | <Q1.1c> = <book> {[<conjunction>] <book>} [<vperfect>] <vpassive> <what_author> <verb_write> {[<conjunction>] |
| 4 | <Q1.1d> = [<time_phrase>] [","] <book> {[<conjunction>] <book>} [<vperfect>] <vpassive> <what_author> <verb_write> "?" |
| 5 | <Q1.2a> = [<interrogative3>] <creator> [<possessive>] <book> {[<conjunction>] <book>} <verb_be> <author> [<interrogative2>] "?" |
| 6 | <Q1.2b> = [<interrogative3>] <author> <verb_be> <creator> [<possessive>] <book> {[<conjunction>] <book>} [<interrogative2>] "?" |
| 7 | <Q1.2c> = <author> [<interrogative3>] <verb_be> <creator> [<possessive>] <book> {[<conjunction>] <book>} [<interrogative2>] "?" |
| 8 | <Q1.3a> = [<interrogative3>] <author> [<vperfect>] [<interrogative1>] <verb_write> <book> {[<conjunction>] <book>} [<time_phrase>] [<interrogative2>] "?" |
| 9 | <Q1.3b> = [<time_phrase>] [","] [<interrogative3>] <book> {[<conjunction>] <book>} [<vperfect>] <vpassive> <author> <verb_write> [<interrrogative2>]"?" |
| 10 | <Q1.4a> = <author> [<vperfect>] [<interrogative1>] <verb_write> <book> {[<conjunction>] <book>} [<prep_time>] < what_time > "?" |
| 11 | <Q1.4b> ::= <book> {[<conjunction>] <book>} [<vperfect>] <vpassive> <author> <verb_write> [<prep_time>] <what_time> "?" |
| 12 | <Q2.1a> = <what_publisher> [<vperfect>] [<interrogative1>] <verb_publish> <book> {[<conjunction>] <book>} [<time_phrase>] "?" |

| 13 | <Q2.1b> = [<time_phrase>] [","] <what_publisher> [<vperfect>] [<interrogative1>] <verb_publish> <book> [[<conjunction>] <book>] "?" |
|----|---|
| 14 | <Q2.1c> = <book> [[<conjunction>] <book>] [<vperfect>] <vpassive> <what_publisher> <verb_publish> [<time_phrase>] "?" |
| 15 | <Q2.1d> = [<time_phrase>] [","] <book> [[<conjunction>] <book>] [<vperfect>] <vpassive> <what_publisher> <verb_publish> "?" |
| 16 | <Q2.2a> = [<interrogative3>] <publisher> [<vperfect>] [<interrogative1>] <verb_publish> <book> [[<conjunction>] <book>] [<time_phrase>] [<interrogative2>] "?" |
| 17 | <Q2.2b> = [<time_phrase>] [","] [<interrogative3>] <publisher> [<vperfect>] [<interrogative1>] <verb_publish> <book> [[<conjunction>] <book>] [<interrogative2>] "?" |
| 18 | <Q2.2c> = [<interrogative3>] <book> [[<conjunction>] <book>] [<vperfect>] <vpassive> <publisher> <verb_publish> [<time_phrase>] [<interrogative2>] "?" |
| 19 | <Q2.2d> = [<time_phrase>] [","] [<interrogative3>] <book> [[<conjunction>] <book>] [<vperfect>] <vpassive> <publisher> <verb_publish> [<interrogative2>] "?" |
| 20 | <Q2.3a> = <publisher> [<vperfect>] [<interrogative1>] <verb_publish> <book> [[<conjunction>] <book>] [<prep_time>] <what_time> "?" |
| 21 | <Q2.3b> = [<prep_time>] <what_time> <publisher> [<vperfect>] [<interrogative1>] <verb_publish> <book> [[<conjunction>] <book>] "?" |
| 22 | <Q2.3c> = <book> [[<conjunction>] <book>] [<vperfect>] <vpassive> <publisher> <verb_publish> [<prep_time>] <what_time> "?" |
| 23 | <Q2.3d> = [<prep_time>] <what_time> <book> [[<conjunction>] <book>] [<vperfect>] <vpassive> <publisher> <verb_publish> "?" |
| 24 | <Q3.1a> = <book> [<of_author>][<by_publisher>][<time_phrase>] <is_of> <what_subject> ? |
| 25 | <Q3.1b> = [<time_phrase>] [,] <book> [<of_author>] [<by_publisher>] <is_of> <what_subject> ? |
| 26 | <Q3.1c> = <field> <possessive> <book> [<of_author>] [<by_publisher>] [<time_phrase>] <interrogative4> ? |
| 27 | <Q3.1d> = [<time_phrase>] [,] <field> <possessive> <book> [<of_author>] [<by_publisher>] <interrogative4> ? |
| 28 | <Q3.2a> = <book> [<of_author>] [<by_publisher>] [<time_phrase>] [<interrogative1>] <is_of> <subject> [<interrogative2>] ? |
| 29 | <Q3.2b> = [<time_phrase>] [,] <book> [<of_author>] [<by_publisher>] [<interrogative1>] <is_of> <subject> [<interrogative2>] ? |
| 30 | <Q3.2c> = <book> [<of_author>] [<by_publisher>] [<time_phrase>] [<interrogative3>] <verb_be> <book_type> <is_of> <subject> [<interrogative2>] ? |
| 31 | <Q3.2d> = [<time_phrase>] [,] <book> [<of_author>] [<by_publisher>] [<interrogative3>] <verb_be> <book_type> <is_of> <subject> [<interrogative2>] ? |
| 32 | <Q3.3a> = [<time_phrase>] [,] <author> [<vperfect>] [interrogative1] <verb_write> [<plural>] <book_type> <verb_have> <what_subject> ? |
| 33 | <Q3.3b> = <author> [<vperfect>] [interrogative1] <verb_write> [<plural>] <book_type> <verb_have> <what_subject> [<time_phrase>]? |
| 34 | <Q3.3c> = [<time_phrase>] [,] <author> [<vperfect>] [interrogative1] <verb_write> [<plural>] <book_type> <is_of> |

| 35 | <what_subject> ? |
|----|---|
| | <Q3.3d> = [<author>] [<vperfect>] [interrogative1] <verb_write> [<plural>] <book_type> <is_of> <what_subject> [<time_phrase>] ? |
| 36 | <Q3.4a> = <publisher> [<vperfect>] [<interrogative1>] <verb_publish> [<plural>] <verb_have> <what_subject> [<time_phrase>] ? |
| 37 | <Q3.4b> = <publisher> [<vperfect>] [<interrogative1>] <verb_publish> [<plural>] <verb_have> <what_subject> ? |
| 38 | <Q3.4c> = <publisher> [<vperfect>] [<interrogative1>] <verb_publish> [<plural>] <is_of> <what_subject> [<time_phrase>] ? |
| 39 | <Q3.4d> = [<time_phrase>] <publisher> [<vperfect>] [<interrogative1>] <verb_publish> [<plural>] <is_of> <what_subject> ? |
| 40 | <Q4.1a> = [<plural>] [book_type] [<verb_have> <subject>] [<by_author>] [<time_phrase>] <interrogative4> ? |
| 41 | <Q4.1b> = [<time_phrase>] [,] [plural][book_type] [<verb_have><subject>] [<by_author>] [interrogative4] ? |
| 42 | <Q4.1c> = [plural][book_type] [<is_of><subject>] [<by_author>] [<time_phrase>] <interrogative4> ? |
| 43 | <Q4.1d> = [<time_phrase>] [,] [plural][book_type] [<is_of><subject>] [<by_author>] <interrogative4> ? |
| 44 | <Q4.2a> = [plural] <book_type> [<verb_have> <subject>] <by_publisher> [<time_phrase>] <interrogative4> ? |
| 45 | <Q4.2b> = [<time_phrase>][,][plural]<book_type> [<verb_have> <subject>] <by_publisher> <interrogative4> ? |
| 46 | <Q4.2c> = [plural]<book_type> [<is_of><subject>] <by_publisher> [<time_phrase>] <interrogative4> ? |
| 47 | <Q4.2d> = [<time_phrase>] [,] [plural] <book_type> <is_of> <subject> <by_publisher> <interrogative4> ? |
| 48 | <Q5.1a> = <book> [<vperfect>] <vpassive> [<publisher>] <verb_publish> <what_place> [<time_phrase>] "?" |
| 49 | <Q5.1b> = [<time_phrase>] [","] <book> [<vperfect>] <vpassive> <verb_publish> <what_place> "?" |
| 50 | <Q5.2> = <publisher><verb_locate><what_place> "?" |
| 51 | <Q6.1a> = [<verb_buy>] <book> <verb_cost> "?" |
| 52 | <Q6.1b> = [<price>] [<possessive>] <book> [<what_price>] "?" |
| 53 | <Q7.1> = <how_many> <book> <in_elib> "?" |
| 54 | <Q7.2a> = <author> [<vperfect>] [<interrogative1>] <verb_write> <how_many> <book> [<time_phrase>] "?" |
| 55 | <Q7.2b> = [<time_phrase>] [","] <author> [<vperfect>] [<interrogative1>] <verb_write> <how_many> <book> "?" |
| 56 | <Q7.3a> = <publisher> [<vperfect>] [<interrogative1>] <verb_publish> <how_many> <book> [<time_phrase>] "?" |
| 57 | <Q7.3b> = [<time_phrase>] [","] <publisher> [<vperfect>] [<interrogative1>] <verb_publish> <how_many> <book> "?" |

This framework also allows adding new syntactic rules which are implemented appropriate treatments.

## IV. Semantic Transformation

After analyzing the syntax of the query, the next step is transforming the syntactic structure to its semantic representation. The semantic representations of queries are based on the semantic model which we have built to represent semantic content of queries.

### A. Semantic model

In semantic model, the verb plays a central role and nouns modify the meaning for it. Relationships are also defined from sub-categories containing verbs, noun phrases, adverbial phrases and prepositional phrases.

For instance, in the case of the verb "viết" ("to write"): "author" is its subject, the relationship is called as « rel_sub »; "book" is its object, the relationship is called as "rel_obj"; APT is the time that the verb "viết" ("to write") is considered, the relationship is called as "rel_time" and it can be multiple values (before, in, after), so we mark with three single values: rel_time1 (before), rel_time2 (in) and rel_time3 (after).

In notation, the convention of the semantic model: if we wish to ask a certain component of BNF query, we'll have to place the question mark ("?") right after it.

From S1_BNF, the semantic model is defined as following:

- *S1_SEM:=(verb_write? ((author, rel_sub), (book0, rel_obj), (APT, rel_time2)))*

In BNF, the elements with "what" labels are those which need to be asked, and they will be marked by a question mark after their name in semantic model. In the case of the elements without "what" labels will belong to Yes/No questions. These questions can also be recognized by identifying used interrogative words.

Another example as following:

*S2:=Nhà xuất bản nào đã xuất bản sách B trong năm 2009?*

*(S2:= which publisher has published book B in 2009?)*

*S2_BNF:=<what_publisher>[<vperfect>]<verb_publish><book>[<APT>] "?"*

In there:

- *what_publisher: Nhà xuất bản nào*

- *vperfect: đã*

- *verb_publish: xuất bản*

- *book: sách B*

- *APT: trong năm 2009*

The semantic model S2_SEM involving to S2_BNF:

- *S2_SEM:=(verb_publish((publisher?, rel_sub), (book, rel_obj), (APT, rel_time2)))*

In BNF, to identify what subject or object is depends on the main verb meaning: if the main verb is "viết" ("to write"), the subject will be "author" and the object will be "book". If the main verb is "xuất bản" ("to publish"), the subject and the object will be "publisher" and "book", …

The transferring from syntactic structure to semantic representation could be processed automatically by the predefined rules. Semantic model helps to eliminate unnecessary components in queries (interrogative words such as: *interrogative1,…, interrogative4*) and remain the key information in presenting the query.

### B. Predefined semantic structures

The full list of semantic structures included in VLPQ framework version 1.0 as follows:

TABLE 2.    Semantic Structures

| Syntactic structure | Semantic structures |
|---|---|
| Q1.1 | (verb_write ((author?, rel_sub), (book, rel_obj), [(year, rel_time2)])) |
| Q1.2 | (verb_be? ((author, rel_sub), ((verb_possessive ((author, rel_sub), (book, rel_obj))), rel_obj))) |
| Q1.3 | (verb_write? ((author, rel_sub), (book, rel_obj), [(time_phrase, rel_time)])) |
| Q1.4 | (verb_write? ((author, rel_sub), (book, rel_obj), [(year?, rel_time2)])) |
| Q2.1 | (verb_publish ((publisher?, rel_sub), (book, rel_obj), [(year, rel_time2)])) |
| Q2.2 | (verb_publish? ((publisher, rel_sub), (book, rel_obj), [(time_phrase, rel_time)])) |
| Q2.3 | (verb_publish ((publisher, rel_sub), (book, rel_obj), (year?, rel_time2))) |
| Q3.1 | (is_of ((is_of (((is_of (book, rel_sub), ([publisher], rel_obj), [(year, rel_time2)])), rel_sub), ([author], rel_obj))), (subject?, rel_obj))) |
| Q3.2 | (is_of? ((is_of (((is_of (book, rel_sub), ([publisher], rel_obj), [(year, rel_time2)])), rel_sub), ([author], rel_obj))), (subject, rel_obj))) |
| Q3.3 | (is_of ((is_of ((book, rel_sub), (author, rel_obj), [(year, rel_time2)])), rel_sub), (subject?, rel_obj))) |
| Q3.4 | (is_of ((is_of ((book, rel_sub), (publisher, rel_obj), [(year, rel_time2)])), rel_sub), (subject?, rel_obj))) |
| Q4.1 | (verb_write ((author, rel_sub), ((is_of(book?, rel_sub), ([subject], rel_obj)), rel_obj), [(time_phrase, rel_time)])) |
| Q4.2 | (verb_publish ((publisher, rel_sub), ((is_of(book?, rel_sub), ([subject], rel_obj)), rel_obj), [(time_phrase, rel_time)])) |
| Q5.1 | (verb_publish (([publisher], rel_sub), (book, rel_obj), [(year, rel_time2)], (location?, rel_loc))) |
| Q5.2 | (verb_locate ((publisher, rel_sub), (location?, rel_obj))) |
| Q6.1 | (verb_cost ((book, rel_sub), (price?, rel_obj))) |
| Q7.1 | (verb_have ((source, rel_sub), (book, rel_obj), (book_amount?, rel_amount))) |
| Q7.2 | (verb_write ((author, rel_sub), (book, rel_obj), [(time_phrase, rel_time)], (book_amount?, rel_amount))) |
| Q7.3 | (verb_publish ((publisher, rel_sub), (book, rel_obj), [(time_phrase, rel_time)], (book_amount?, rel_amount))) |

Respectively, each syntactic structure is represented by a syntactic rule, a semantic structure is defined.

## V. CONCLUSION

Building computer systems with ability of understanding human's natural language is a challenging research. Only pure syntax analyzing does not let computer understand human language. We have proposed the semantic representation model to process Vietnamese query forms in determined application domains. Some gained results show that this is a right and promising approach, due to the lacking of methods that help computer to understand all terms presented by human language at the present.

In VLQP framework, the semantic model is an original feature we have addressed. This semantic model contributes to the syntax analyzing and representation of Vietnamese query forms involving to application domain. We also propose transforming rules to transform syntactic structures to their semantic representation.

The framework has been deployed and tested with 200 Vietnamese queries. Results of manual testing stage show that the framework meets all of described requirements. This framework can be further developed to work with more new forms of Vietnamese queries. From this model framework, we anticipate building more frameworks to handle Vietnamese queries for other application domains.